\documentclass[11pt]{article}
\usepackage{amsmath}
\usepackage{cite}
\title{Algebraic entropy for semi-discrete   equations} 

\author{D.K. Demskoi \\ School of Computing and Mathematics \\ Charles
  Sturt University\\ 
  Locked Bag 588, Wagga Wagga, NSW 2678, Australia
  \and C-M. Viallet \\ LPTHE, Universit\'e Pierre et Marie
  Curie\footnote{Sorbonne Universit\'es} \\ Centre National de la
  Recherche Scientifique, UMR 7589 \\ Bo\^ite 126, 4 Place Jussieu,
  75252 Paris Cedex 05, France}

\begin{document}
\maketitle

\begin{abstract}
We extend the definition of algebraic entropy to semi-discrete
(difference-differential) equations. Calculating the entropy for a
number of integrable and non integrable systems, we show that its
vanishing is a characteristic feature of integrability for this type
of equations.
\end{abstract}

\section{Introduction}
Algebraic entropy~\cite{BeVi99} was introduced as a measure of the
complexity of discrete systems, originally for maps over finite
dimensional spaces and finite order discrete (difference)
equations. It relates to general notions of complexity of
maps~\cite{Yo87,Ar90,Ve92,FaVi93,RuSh97,AbHaHe00,DiFa01,Gr03,Ha05,Fr06,Mc07,Si07}.
Its use as an integrability detector for such systems is by now firmly
established~\cite{HiVi98,HiVi97b,Ta01b,GrHaRaVi09}. Its definition has
been further extended to infinite-dimensional systems, that is to say
partial difference equations~\cite{TrGrRa01,Vi06}, and it provides a very
good characterisation of integrability for these systems as
well~\cite{Vi08}. In all cases integrability causes a drop in
complexity, making the entropy vanish, while the entropy is nonzero for
generic non integrable systems.

We give in Section~\ref{definition} a generalisation of the concept to
systems which mix difference and differential equations.  These
semi-discrete equations constantly appear, for example from partial
discretisation of continuous systems or in the description of
symmetries of integrable equations, and the "low growth requirement"
has already been used for such systems, in conjunction with
singularity confinement in the earlier Ref.~\cite{TaRaGrOh99}.  In
Section~\ref{singularity} we recall the link between the singularity
structure and the value of the entropy. In Section~\ref{calculation}
we describe how to calculate the entropy.

In Section~\ref{nls} we calculate the entropy for two discretisations
of the nonlinear Schr\"odinger equation: the integrable one given by
the Ablowitz-Ladik system~\cite{AbLa76} and a more naive non
integrable one.  In Section~\ref{ydkn} we calculate the entropy of
Yamilov's form~\cite{Ya83} of the discrete Krichever-Novikov equation
(see also Ref.~\cite{Ad98}). In Section~\ref{nonautonomous} we apply
the definition to nonautonomous systems, the master symmetries of the
Volterra, modified Volterra, and discrete Calogero-Degasperis
equations given in Ref.~\cite{ChYa95}.  

The outcome of all our calculations is that a vanishing entropy is,
here again, a characteristic  feature of integrability.

We conclude by indicating some directions for further studies.

\section{Definition}
\label{definition}
We deal with difference-differential equations of the form 
\begin{equation}
\label{recurrence}
u_{n+1} (t) = \frac{A  +  B\;  u_{n-k}(t)}{ C  +  D\; u_{n-k}(t) },
\end{equation}
where $t$ is a $\nu$-tuple of ``times'' $[t_1, t_2, \dots, t_\nu]$ and
$A,B,C,D$ are rational functions of $u_n(t)$, $\dots$, $u_{n-k+1}(t)$
and a finite number of their derivatives $\partial_{t_q} u_n,
\partial_{t_q t_r}^2 u_n,\dots,\partial_{t_q} u_{n-k+1}, \partial_{t_q
  t_r}^2 u_{n-k+1},\dots$, with respect to $t_1, t_2, \dots,
t_\nu$. The functions $A,B,C,D$ may explicitly depend on the times
$t_1, t_2, \dots, t_\nu$ and on the discrete index $n$ (non autonomous
case).

Such equations define invertible recurrences of order $k+1$, that is to say 
a very large number of interesting equations.

The first step is to define from Eq.~(\ref{recurrence}) a homogeneous map
by projectivisation. The simplest presentation of this process is to
write

\begin{eqnarray*}
\left[
\begin{matrix}
u_n &= &X_1/X_{k+1} \\
u_{n-1} & =&  X_2/X_{k+1} \\
& \dots & \nonumber \\
u_{n-k+1} &=& X_k/X_{k+1} 
\end{matrix}
\right]
\qquad  \mbox { and } \qquad
\left[ 
\begin{matrix}
u_{n+1} &= &Y_1/Y_{k+1} \\
u_{n} & =&  Y_2/Y_{k+1} \\
& \dots & \\
u_{n-k+2} &=& Y_k/Y_{k+1}
\end{matrix}
\right].
\end{eqnarray*}

Equation (\ref{recurrence}) defines a map from $(u_n, u_{n-1}, \dots,
u_{n-k+1})$ to $(u_{n+1}, u_{n}, \dots, u_{n-k+2})$ which translates
into a map $\varphi$ from the $(k+1)$-tuple $ x = [X_1, X_2, \dots ,
  X_{k+1}]$ to its image $ y = [Y_1, Y_2, \dots , Y_{k+1}]$. All
components of $y$ are polynomials in $X_1, X_2, \dots , X_{k+1}$ and a
finite number of their derivatives with respect to the times $t_1,
t_2, \dots, t_\nu$. If we attribute to all $X$'s and any of their
derivatives a weight 1, and weight $0$ to all coefficients, even if
they depend on the times or the discrete indices, these polynomials
are homogeneous with some weight $w_1$.

We then iterate the map $\varphi$.  Since we have homogeneised the
coordinates, any factor common to the $k+1$ components should be
factored out. Doing so we generate a sequence of $(k+1)$-tuples, all with
a well defined weight (using the simple fact that derivatives with
respect to the times satisfy Leibnitz' rule). This defines a sequence of
weights $w_1, w_2, \dots$.

The sequence $\{w_m\}$ satisfies the property
\begin{equation}
\label{subadditivity}
w_{l+m} \leq \;  w_l \cdot w_m.
\end{equation}

{\bf Definition:} The entropy is  defined as
\begin{equation}
\label{entropy}
\epsilon = \lim_{m \rightarrow \infty} \frac{1}{m} \log( w_m).
\end{equation}

{\bf Claim:} The previous property Eq.~(\ref{subadditivity}) of the
sequence of weights ensures the existence of this limit.

\section{Singularity structure and entropy}
\label{singularity}

There is a deep link between the singularity structure~\cite{RaGrTa92}
of the evolution and the entropy. The defining equation
(\ref{recurrence}) is such that the map $\varphi$ has a polynomial
inverse $\psi$. This allows us to define two polynomials
$\kappa{_\varphi}$ and $\kappa{_\psi}$ as in Ref.~\cite{BeVi99}: since
$\varphi$ and $\psi$ are inverse of each other, their product appears
as a mere multiplication of the components by some polynomial
\begin{eqnarray*}
(\psi \cdot \varphi)(x) = \kappa_\varphi(x). id (x)\; \mbox{ and } \;
(\varphi \cdot \psi)(x) = \kappa_\psi(x). id (x),
\end{eqnarray*}
where $id$ is the identity map.

During the iteration of $\varphi$, the appearance of some factor $F$
common to all components means that the hypersurface given by $F=0$ is
sent by the corresponding iterate of $\varphi$ on a singular set of
$\varphi$ (singularity means that the polynomial image is $[0,0,
  \dots, 0]$). The only possible factors are thus the factors
composing $\kappa_\varphi$ and their proper images by $\psi$. We will
give some examples of this in the sequel.

If there is no factorisation, the sequence of weights is purely exponential
and 
\begin{eqnarray*}
w_n = w_1^n, \qquad \epsilon= \log( w_1).
\end{eqnarray*}

{\bf Claim:} Any drop in degree during the iteration comes from the
possible presence of common factors, and is related to the singularity
structure for the geometrical reason explained above.

We will illustrate this fundamental property in some of the examples
studied below.

\section{Actual calculations}
\label{calculation}

One may of course calculate the iterates of $\varphi$. The main
obstacle is the size of the explicit expressions, which rapidly
exceeds the capacity of hand calculations as well as computer-aided
ones.

The first issue is that we work with an algebra generated by an
infinite-dimensional set: the $X$'s and their derivatives.  One should
however notice that, at every finite order, one uses only a finite
subset of the generators, and manipulates only polynomials. This makes
the calculation of a few of the first terms of the sequence $\{w_m\}$
possible.

To go further we will use the method described in Ref.~\cite{Vi08b}. We
calculate the iterates of $\varphi$ starting with a rational initial
condition, of the form $\left[ X_1(t_1, t_2, \dots, t_\nu), \dots ,
  X_{k+1}(t_1, t_2, \dots, t_\nu)\right]$, with all the $X$'s
polynomials in the times appearing in Eq.~(\ref{recurrence}).

 One then produces the sequence of degrees $\{d_n\}$ of the
 iterates. The point is that,{ \em for generic initial conditions,
   this sequence retains the feature of the sequence of weights
   $\{w_m\}$ we are interested in: its asymptotic behaviour}. In some
 cases, and with a judicious choice of initial conditions, the two
 sequences $\{d_n\}$ and $\{w_m\}$ {\em may coincide exactly}.

The next step is to extract the value of the entropy, which is an
asymptotic quantity, from the first few terms of the sequence of
weights (resp. degrees). The standard heuristic method is to fit a
generating function of the sequences with rational fractions.
\begin{eqnarray}
\label{generating}
g_w(s) = \sum_{n=0}^\infty w_n \; s^n = \frac{P_w(s)}{Q_w(s)}, \qquad
\mbox{or} \qquad g_d= \sum_{n=0}^\infty d_n \; s^n =
\frac{P_d(s)}{Q_d(s)}.
\end{eqnarray}
The method has already been shown to work remarkably well for maps and
lattice equations \cite{Vi08,GrHaRaVi09}, and leads to extremely simple rational fractions
with integer coefficients. This is related to the algebraic properties
of the entropy, in particular the still conjectured fact that the entropy
always is the logarithm of an algebraic integer~\cite{BeVi99,HaPr05},
but we will not dwell on that here.

Once the first terms of the sequence are fitted with the Taylor
coefficients of a rational fraction, the method becomes predictive,
and is a good check of the validity of the generating function. The
location of the smallest pole of the denominator ($Q_d(s)$ or
$Q_w(s)$) of the generating function gives the exact value of the
entropy.

{\bf Remark 1:} When the entropy vanishes, the growth of the sequences is
polynomial. A quick test is to calculate the successive discrete
derivatives of the sequences.  Linear growth is equivalent to a bounded
first derivative, quadratic growth is equivalent to a bounded second
derivative, and so on.

{\bf Remark 2:} The exact value of the entropy may sometimes be
extracted from the singularity analysis. In one of the cases we will
be able to do that (see Section~\ref{naive-nls}).

{\bf Remark 3:} It is quite possible to perform the calculation of the
iterates without going to the homogeneous description. The expressions
are then rational fractions rather than polynomials, and the degree to
retain is the maximum degree of their numerators and denominators. The
sequence of degrees one produces is usually different from the one
given by the homogeneous form. It nevertheless yields the same value
of the entropy, since the discrepancy between the two sequences is at
most a multiplicative factor smaller than or equal to the order $k+1$.

\section{Two semi-discrete  non linear Schr\"odinger equations}
\label{nls}

One source of semi-discrete equations is the incomplete discretisation
of partial differential equations. The nonlinear Schr\"odinger
equation is a typical example of such systems:
\begin{eqnarray*}
\label{contiuous_nls}
i u_t + u_{xx} + 2 \vert u \vert^2 u = 0,
\end{eqnarray*}
with $u$ a complex function of a space variable $x$ and a time
variable $t$.  If only the spatial coordinate $x$ is discretised, we
get a semi-discrete equation with one continuous time variable
($\nu=1$). We will examine two cases. The first is the integrable one
given by Ablowitz and Ladik~\cite{AbLa76}:
\begin{eqnarray}
 \label{abla}
i\;  \partial_t u_j + (u_{j-1} - 2 u_j + u_{j+1}) / h^2 +  \vert
u_j\vert^2 (u_{j-1}+u_{j+1}) = 0.
\end{eqnarray}
The second one is a slightly more naive one
\begin{eqnarray}
\label{naive}
i \; \partial_t u_j + (u_{j-1} - 2 u_j + u_{j+1}) / h^2 + 2 \vert
u_j\vert^2 u_j =0.
\end{eqnarray}
Here $h$ is the discretisation step.  Both cases are perfectly
sensible as discretisations of Eq.~(\ref{contiuous_nls}). However they
have been shown to behave very differently~\cite{HeAb89},
the reason being that only the form Eq.~(\ref{abla}) retains one of the
main features of the original equation: its integrability. We will
evaluate their respective entropies.

\subsection{Ablowitz-Ladik system}
\label{ablowitz-ladik}
We rewrite equation (\ref{abla}) as a system for the real and
imaginary parts $q_j$ and $r_j$ of $u_j$, with some rescaling:
\begin{eqnarray}
 \partial_t {q_n} = q_{n+1} - 2\; q_n + q_{n-1} + q_n \; r_n ( q_{n+1} + q_{n-1}) \nonumber \\
- \partial_t{r_n} =  r_{n+1} - 2\; r_n + r_{n-1} + q_n \; r_n ( r_{n+1} + r_{n-1}).
\end{eqnarray}
The homogenisation proceeds by setting
\begin{eqnarray*}
q_n = \frac{X}{V}, \qquad r_n=\frac{Y}{V}, \qquad q_{n-1}=\frac{Z}{V},
\qquad r_{n-1} = \frac{U}{V}.
\end{eqnarray*}
The map $\varphi$ in the homogeneous coordinates $[ X, Y, Z, U, V ]$ reads
\begin{eqnarray*}
\varphi: 
\begin{pmatrix} 
 X\cr Y\cr Z\cr U\cr V 
\end{pmatrix}
\longrightarrow
\begin{pmatrix} 
V^2 ( X' +2\, X -Z) - X ( V V' + Y Z) \cr
 V^2 (- Y' +2\, Y -U) + Y  ( V V' - X U) \cr
X ( V^2 + X Y) \cr
Y ( V^2 + X Y) \cr
V ( V^2 + X Y)
\end{pmatrix}
\end{eqnarray*}
which is a degree $3$ birational map over the algebra generated by 
\begin{eqnarray*}
X, X', X'', \dots, Y, Y', Y'', \dots Z,Z',Z'', \dots, V,V',V'' \dots
\end{eqnarray*}
where $X'= \partial_t X$, etc.

The inverse is
\begin{eqnarray*}
\psi: 
\begin{pmatrix} 
 X\cr Y\cr Z\cr U\cr V 
\end{pmatrix}
\longrightarrow
\begin{pmatrix} 
Z(V^2+UZ) \cr
U(V^2+UZ) \cr
V^2( Z' +2\, Z -X) -Z( V V' + U X )\cr
V^2( -U' +2\, U -Y) +U( V V' - Y Z) \cr
V(V^2+ UZ)
\end{pmatrix}.
\end{eqnarray*}
The direct calculation of the iterates of $\varphi$ produces 
\begin{eqnarray*}
\{ w_n \} =  1,\quad 3, \quad 9,\quad 19
\end{eqnarray*}
which is not long enough to guess the asymptotic behaviour.

The specialisation of the initial condition with linear functions of
$t$, which provides the simplest nontrivial rational initial values for
$q_n$, $r_n$, produces the sequence
\begin{eqnarray*}
\{ d_n \} = 1, 3, 9, 19, 33, 51, 73, 99, 129, 163, 201, 243, 289,
\dots
\end{eqnarray*}
The sequence is (redundantly) fitted by the generating function
\begin{eqnarray*}
g(s) = {\frac {1+3\,{s}^{2}}{ \left(1- s \right) ^{3}}},
\end{eqnarray*}
showing quadratic growth, vanishing entropy, i.e. integrability.

Consider degree drop and singularity pattern. The first drop in degree
comes with $\varphi^3$, since $3 \times 9 - 19 = 8$.  We know that
this happens when $\varphi$ hits some singular point. The pattern is
the following: the surface $\Sigma_+ : V^2+X Y =0$ is sent (blown
down) to $[- X ^{2}, V ^{2},0,0,0]$ by $\varphi$. At the next step it
goes to $[0,0,X^2,-V^2,0]$, a singular point which $\varphi$ blows up
to the surface $\Sigma_- :V^2+UZ=0$. The third iterate sends
``smoothly'' $\Sigma_+ $ into $\Sigma_-$, and the drop of the degree
is due to the presence of the factor $\Sigma_+^4$.

\subsection{Naive discretisation of the nonlinear Schr\"odinger equation}
\label{naive-nls}
Consider now equation (\ref{naive}) in terms of the same variables $q_n$, $r_n$:
\begin{eqnarray}
 \partial_t{q_n} = q_{n+1} - 2\; q_n + q_{n-1} + 2  q_n^2 \; r_n \nonumber 
\\
- \partial_t{r_n} =  r_{n+1} - 2\; r_n + r_{n-1} + 2 q_n \; r_n^2 .
\end{eqnarray}
With the same notations as in the previous case, the map $\varphi$  reads
\begin{eqnarray*}
\varphi: 
\begin{pmatrix}  X\cr Y\cr Z\cr U\cr V \end{pmatrix}
 \longrightarrow
\begin{pmatrix} 
Z V^2\cr
U V^2 \cr
 \cr
( 2 Z -X) V^2    - 2 Z^2 U + Z' V^2 - Z V V' \cr
 ( 2 U -Y ) V^2  - 2 Z U^2 - U' V^2 + U V V' \cr
V^3
\end{pmatrix}.
\end{eqnarray*}
The sequence of degrees we get is purely exponential
\begin{eqnarray*}
\{ d_n\} = 1, 3, 9, 27, 81, 243,\dots
\end{eqnarray*}
It is straightforward to prove that the sequence of weights
is exactly the same; that is, there is no degree drop, $ w_n =
3^n$.

The inverse of $\psi$ is related to $\varphi$ by a linear
similarity. If $\lambda$ is the permutation
\begin{eqnarray*}
\lambda: [X,Y,Z,U,V] \longrightarrow [Z,U,X,Y,V], 
\end{eqnarray*}
then
\begin{eqnarray*}
\psi = \lambda \cdot \varphi \cdot \lambda.
\end{eqnarray*}
The polynomials $\kappa_\varphi$ and $   \kappa_\psi$ are just
\begin{eqnarray*}
\kappa_\varphi = \kappa_\psi = V^8.
\end{eqnarray*}
The ``hypersurface'' ${V=0}$ is sent by $\varphi$ to one of its fixed
points $ [0,0,Z,U,0]$. Consequently the later iterates cannot hit any
singular point, and there will be no factorisation. This proves that
the entropy is exactly $ \epsilon = \log (3)$ in contrast with the
vanishing entropy of the integrable discretisation of Ref.~\cite{AbLa76}.

\section{Generalised symmetries of integrable equations}
\label{ydkn}

A source of integrable semi-discrete equations is the symmetry
approach to integrability of discrete
systems~\cite{MiShYa87,AdShYa00,Ya06,LeWiYa11,MiWaXe11,MiWaXe11b,MiWa11}.
A fundamental equation is Yamilov's form of the discrete
Krichever-Novikov equation~\cite{Ya83}.  The equation is
\begin{eqnarray}
\partial_t u_{n}  =\frac {R\,(u_{n+1}, u_n, u_{n-1})}  {u_{n+1}-u_{n-1}},
\end{eqnarray}
with six free parameters $\beta_i, \quad i=1\dots 6$:
\begin{eqnarray}
\hskip -.5truecm 
R(u_{n+1}, u_n, u_{n-1}) & = & \left( \beta_1 \,{u_{n}}^{2}+2\, \beta_2 \,u_{n}+
\beta_3 \right) u_{n+1} \; u_{n-1} + \left(\beta_2\,{u_{n}}^{2}+ \beta_4\,u_{n}+
\beta_5 \right) \left( u_{n+1}+ u_{n-1} \right)
\nonumber 
\\
&  &+ \beta_3\,{u_{n}}^{2}+2\, \beta_5\,u_{n}+ \beta_6.
\label{defR}
\end{eqnarray}
Using an obvious notation, the homogeneous version of the corresponding map reads
\begin{eqnarray*} 
%\hskip -.7truecm 
\varphi: 
\begin{pmatrix} X \cr Y \cr Z \end{pmatrix} \longrightarrow
\begin{pmatrix}
-Z \left[
{\beta_2 X}^{2}Y+ \beta_3 {X}^{2}{Z} +\beta_4 XY{Z}+ \beta_5 {Z}^{2}
 \left( Y+2\,X \right) + \beta_{6}{Z}^{3} + Y  ( X' Z - X  Z') \right]
\cr
X \left[ \beta_1 {X}^{2}Y + \beta_2 {X}Z \left( 2\,Y+X \right) +  \beta_3 Y{Z}^{2} + 
\beta_4 {X}{Z}^{2} + \beta_5 {Z}^{3} - Z ( X' Z - X Z' ) \right]
 \cr
Z \left[
\beta_1 {X}^{2}Y + \beta_2 {Z}X \left( 2\,Y+X \right)  + \beta_3 Y{Z}^{2} + 
\beta_4 X{Z}^{2} + \beta_{5}{Z}^{3} -Z ( X' Z - X Z') \right]
\end{pmatrix},
\end{eqnarray*}
The sequence of degrees obtained with degree 1 initial conditions is:
\begin{eqnarray*}
\{d_n \} = 1 , \; 4, \; 10, \; 20, \; 34, \; 52, \; 74, \;
100, \; 130, \; 164, \; 202, \; 244, \; 290, \; 340,
\; 394, \; 452, \dots
\end{eqnarray*}
fitted by the generating function
\begin{eqnarray*}
g(s) = {\frac { \left( 1+s \right) \left( 1+{s}^{2} \right) }{
    \left(1- s \right) ^{3}}},
\end{eqnarray*}
meaning quadratic growth and vanishing entropy.

Consider the singularity structure. We have
\begin{eqnarray*}
\kappa_\varphi = A^2 \,  B^3 \,  C,
\end{eqnarray*}
with
\begin{eqnarray*}
A & = & Z, \quad B = \beta_1 X^2 Y+ \beta_2 X Z \left( 2\,Y+X \right)
+ \beta_3 Y {Z}^2 + \beta_4 X Z^2+ \beta_5 Z^{3}+ Z \left( - X' Z+X Z'
\right) \\ C & = & 2 \; X Z \left[ \left( \beta_1 \beta_5 + \beta_2
  \beta_3 - \beta_2 \beta_4 \right) X^2 + \left( \beta_2 \beta_5 +
  \beta_3^2 - \beta_4^2+ \beta_1 \beta_6 \right) X Z + \left ( \beta_2
  \beta_6 - \beta_4 \beta_5 + \beta_3 \beta_5 \right) Z^2 \right]
\\ && + \left( \beta_1 \beta_3 - \beta_2^2\right) X^4 + \left( \beta_3
\beta_6 - \beta_5^2 \right) Z^4 + \left(X Z' - X'Z \right) ^2.
\end{eqnarray*}

For example, at step $\varphi^2$ the product  $A^3 \, B$ factors out
causing a degree drop of $6 = 4 \times 4-10 = 3+3$.

At the next step the factor is $ B^3 \,  C \, \varphi_*(B) $
causing a degree drop of $ 20 = 4 \times 10 -20 = 3 \times 3 + 4 + 7 $
(here $\varphi_*(B)=0$ is the equation of the proper image by $\psi$
of the surface of equation $B=0$, i.e. of the pull-back by
$\varphi$). So the geometrical interpretation of the degree drops is
perfectly in order.

{\bf Remark:} Changing in an arbitrary way the expression of $R$
completely changes the result: for example introducing a new
independent parameter $\beta_7$ as the coefficient of $u_n^2
\,(u_{n-1} + u_{n+1})$ in Eq.~(\ref{defR}) leads to a sequence of
degrees $ 1, 4, 10, 24, 58, 140, 338, 816, 1970,\dots $, fitted by $
g(s) = (1+s)^2/(1-2 s - s^2)$, meaning an entropy of $\log ( 1 +
\sqrt{2})$.

\section{Non autonomous systems}
\label{nonautonomous}

Three interesting semi-discrete equations are provided by master
symmetries of the Volterra, modified Volterra, and  discretised
Calogero-Degasperis equations, respectively Eqs.~(6), (20) and
(26) of Ref.~\cite{ChYa95}.

\begin{itemize}

\item
The first equation is 
\begin{eqnarray}
\partial_t u_n = u_n \left[ (\epsilon + n + 2)\, u_{n+1} + u_n
  -(\epsilon + n - 1)\, u_{n-1} \right].
\end{eqnarray}
It is nonautonomous in the discrete variable $n$, but the coefficient
$\epsilon$ is constant.

The  corresponding map reads
\begin{eqnarray*}
\begin{pmatrix} X \\ Y \\ Z \end{pmatrix}
\longrightarrow
\begin{pmatrix} 
X' Z - X Z' -X^2 + (\epsilon + n -1) \, X Y \\
(\epsilon + n +2) \, X^2 \\
(\epsilon + n +2) \, X Z
\end{pmatrix}
\end{eqnarray*}
where $X'=\partial_t X$, etc...

One finds the sequence of degrees 
\begin{eqnarray*}
\{d_n\}= 1, 2, 4, 8, 13, 20, 28, 38, 49, 62, 76, 92, 108, 127, 146,
166, 188, 211, 236, 262,290,319,350,\dots
\end{eqnarray*}
fitted by the generating function
\begin{eqnarray*}
g(s) = {\frac { \left( 1+2\,{s}^{3} \right) \left(1 -{s}^{12}+
    {s}^{13} \right) }{ \left(1+ s\right) \left(1- s \right) ^{3}}}.
\end{eqnarray*}
We have quadratic growth of the degree, vanishing entropy, and integrability.

\item 
The second equation is given by
\begin{eqnarray}
\label{modif}
\partial_t v_n = ( \lambda - v_n ) \; v_n \; \left[ (\epsilon +n +1) \,
  v_{n+1} - (\epsilon + n - 1) \, v_{n-1} \right] + \lambda \; v_n ( v_n
- \lambda/2).
\end{eqnarray}
Here the parameter $\lambda$ {\em depends on the time} and satisfies
  the condition 
\begin{equation}
\label{lambdadep}
\partial_t \lambda= \lambda^3/2.
\end{equation}
Equation (\ref{modif}) is thus non autonomous both in the discrete and the
continuous directions. 

For this case, the map $\varphi$ is
\begin{eqnarray*}
\begin{pmatrix} X \\ Y \\ Z \end{pmatrix}
\longrightarrow
\begin{pmatrix} 
 2\, Z ( {X'}\,{Z}- X{Z'}) +  2  X Y (\lambda Z - X) ( \epsilon+n-1) 
+ \lambda X Z ( \lambda Z - 2 X) \\
 2\,  X^2 ( \lambda Z - X) (  \epsilon +n + 1)     \\
2\, X Z ( \lambda Z    - X ) ( \epsilon +n + 1)
\end{pmatrix}.
\end{eqnarray*}

We get the sequence 
\begin{eqnarray*}
 \{ d_n \}= 1,3,9,19,33,51,73,99,129,\dots
\end{eqnarray*}
which has quadratic growth: the second derivative is constant (value
$4$).  

Notice that if we do not impose the correct time dependence of
$\lambda$ given by Eq.~(\ref{lambdadep}), supposing that $\lambda$ is
constant, the entropy does not vanish anymore. We get  the
sequence $1,3,9,22,51,116,262,590,1327,\dots$ fitted by the generating
function 
$$ g(s) = \frac{ {1+{s}^{2}-{s}^{4}}}{( 1-s ) ({s}^{3}-{s}^{2}-2\,s+1)}
$$ 
and giving an entropy of approximately
$\log ( 2.246979604 )$ [ logarithm of the inverse of the smallest root
  of $\left( {s}^{3}-{s}^{2 }-2\,s+1 \right)$].

\item
The third equation reads
\begin{eqnarray}
4\;  \partial_t v_n = ( 1 - v_n^2  ) \left[ ( b^2 - a^2 v_n^2) \left( 
\frac{\epsilon + n }{v_{n+1} + v_n} - \frac{\epsilon +n - 1}{ v_n + v_{n-1}} 
\right) + a^2 \, v_n \right],
\end{eqnarray}
with $a=\lambda + \mu $ and $ b = \lambda - \mu$, the two coefficients
$\lambda$ and $\mu$ verifying the same equation (\ref{lambdadep}).

%% map ?
We will not write explicitly the formula for the map $\varphi$ here.

We get the sequence
\begin{eqnarray*}
 \{ d_n \}= 1,4,13,28,49,\dots
\end{eqnarray*}
which has constant second derivative (value $6$).  

If one took $a$ and $b$ constant, forgetting their time dependence,
the sequence of degrees would be $1, 4, 15, 42, 107, 264, 643,\dots$
showing exponential growth, and non vanishing entropy $\log( 1+\sqrt{2})$.

\end{itemize}

\section{Conclusion}

All our calculations, without exception, show that the vanishing of
the entropy is the hallmark of integrability for the extended class
of semi-discrete systems, as it was for the purely discrete ones.

Our study can be complemented in the future in a number of directions:
\begin{itemize}
\item
In order to completely validate the calculations made with degrees
rather than weights, it is sufficient to examine the factors dropping
from the successive images under the iteration, and check that all
factors are in agreement with the geometrical description given in
Section~\ref{singularity}. This will guarantee the absence of
spurious factorisation, due to the possible non-genericity of the
initial conditions.
\item
We have not described here equations with more than one time ($\nu
\geq 2$), as for example the 2-dimensional infinite Toda field theory
\begin{eqnarray*}
\partial^2_{xt}  u_n = \exp\; ( u_{n+1} - 2 \; u_n + u_{n-1})
\end{eqnarray*}
which can be turned into an equation of the form (\ref{recurrence}) by
the change of variable $ \exp\; ( u_n) = v_n$.  Our definition applies
directly to this type of higher dimensional systems, and the result is
the same.
\item
Another direction to explore is lattice equations, that is to say
systems with more than one discrete index, to which the definition can
be adapted.
\item
Finally, the homogeneous setting we have introduced is more general
than the recurrence (\ref{recurrence}) we started from. The definition
we gave can thus be applied to a much wider class of semi-discrete
equations.
\end{itemize}

%\bibliography{../ref}

\end{document}